# Perfect diffraction metagratings supporting bound states in the continuum and exceptional points

Zi-Lan Deng[1,2,*], Feng-Jun Li[1], Huanan Li[2], Xiangping Li[1], Andrea Alù[2,3,†]

[1]*Guangdong Provincial Key Laboratory of Optical Fiber Sensing and Communications, Institute of Photonics Technology, Jinan University, Guangzhou 510632, China.*

[2]*Photonics Initiative, Advanced Science Research Center, City University of New York, New York, NY 10031, USA*

[3]*Physics Program, Graduate Center, City University of New York, New York, NY 10016, USA*

E-mail: [†]aalu@gc.cuny.edu, [*]zilandeng@jnu.edu.cn

**Resonance coupling in non-Hermitian systems can lead to exotic features, such as bound states in the continuum (BICs) and exceptional points (EPs), which have been widely employed to control the propagation and scattering of light. Yet, similar tools to control diffraction and engineering spatial wavefronts have remained elusive. Here, we show that, by operating a suitably tailored metagrating around a BIC and EPs, it is possible to achieve an extreme degree of control over coupling to different diffraction orders in metasurfaces. We stack subwavelength metallic slit arrays on a metal-insulator-metal waveguide, enabling a careful control of the coupling between localized and guided modes. By periodically tuning the coupling strength from weak to strong, we largely tailor the overall spectrum and enable the emergence of singular features, like BICs and EPs. Perfect unitary diffraction efficiency with large spectrum selectivity is achieved around these singular features, with promising applications for arbitrary wavefront shaping combined with filtering and sensing.**

Metasurfaces can manipulate light propagation at will based on arrays of tailored subwavelength inclusions [1-3]. Since they typically operate within the far-field radiation continuum, they are inherently a non-Hermitian (open) platform. Their dispersion and spectral features can be engineered to support bound states in continuum (BICs) [4-6] and exceptional points (EPs) [7-9]. BICs are self-sustained eigenmodes embedded within the radiation continuum, typically manifested by a vanishing linewidth as we continuously vary a parameter associated with the structure symmetry [10-13] or the coupling strength between multiple modes or resonance channels [14-19]. The associated high-Q resonances around a BIC have been widely employed to construct ultra-sharp transmission and reflection spectra, leading to performance-enhanced applications, such as BIC lasing [20, 21] and sensing [22]. The non-Hermitian nature of metasurfaces enables also the emergence of EPs, which are branch point singularities in the *k*-vector space of the metasurface at which multiple eigen-frequencies and the corresponding eigenmodes coalesce [23-25]. Such singularities offer potential opportunities for enhanced sensitivity, due to their super-linear dependence to small perturbations [26-28]. Currently, both BICs and EPs in metasurfaces have been mostly focused on the manipulation of the spectral/temporal properties of the impinging optical signals, namely to control the zero-th diffraction-order transmission/reflection features for incident plane-waves [10-13, 23-25].

In parallel, few-diffraction-order (FDO) metagratings, capable of suppressing undesired diffraction orders and redirect the incident power to preferred directions with unitary efficiency, have been developed an efficient platform for extreme wavefront engineering with less spatial resolution demand [29-40] when compared with conventional phase-gradient metasurfaces [41, 42]. In this paper, we introduce a platform to control the emergence of BICs and EPs near the Brillouin edge of an FDO metagrating, tailoring the diffraction spectra in extreme ways. Specifically, we show how, by continuously tuning the coupling strength between localized and guided resonant modes, we can map the dispersion relation onto the trajectory of perfect diffraction towards the -$1^{st}$ order. Our work connects BICs and EPs with light diffraction, offering novel opportunities for nanophotonic engineering, such as embedded wavefront shaping and spatial wavefront sensing [43-46].

The geometry of interest is shown in Fig. 1(a), and it is composed of a layer of subwavelength metallic slits with thickness $t$, width $w$ and periodicity $p$, a dielectric layer of refractive index $n$, and a metallic substrate. The metallic slit array supports localized modes (LM) within the gap area, while the dielectric layer supports guided modes (GM). Both LMs and GMs can be exploited to perform desired diffraction order management (Supplementary Fig. S1a, b), based on the LM flat dispersion or the folded GM dispersion in the FDO regime, where only $0^{th}$ and $1^{st}$ or -$1^{st}$ diffraction orders are within the light cone, as shown by the green patches in Supplementary Fig. S1a, b. At the LM or GM resonances, the -$1^{st}$ diffraction efficiency in reflection approaches unity, while specular reflection is suppressed, leading to a peak in $R_{-1}$ and a dip in $R_0$ at the trajectory of the LM or GM dispersion curves (Supplementary Fig. S2). By simultaneously combining LM and GM resonances in the FDO regime of a metagrating (inset of Fig. 1a), the coupling between LM and GM offers rich opportunities to engineer the overall scattering response involving multiple diffraction channels.

Generally, the subwavelength slits in our metagrating operate as a periodic perturbation for the waveguide modes in the underlying dielectric region, leading to avoided anti-crossing between forward and backward GM. The coupling strength can be readily tuned by the LM resonance in the gap area, which is controlled by the slit layer thickness $t$. When the LM resonance frequency coincides with the crossing point of the forward/backward GMs ($t=0.425p$), maximum coupling is achieved (Fig. 1b). In this strong coupling scenario, a large Rabi splitting between the GMs (higher and lower quadratic bands) can be observed. Between the quadratic bands, a flat band emerges associated with the LM, with a vanishing linewidth at the Brillouin boundary, corresponding to a BIC. The dispersion of coupled GMs and LM is reflected into the dip and peak trajectories of the specular reflection coefficient $R_0$ (left panel of Fig. 1b) and -1$^{st}$ diffraction $R_{-1}$ (right panel of Fig. 1b), respectively, offering a flexible way to manipulate light diffraction and wavefront shaping based on high-Q BICs.

In contrast, when the LM resonance is tuned far away from the forward/backward GM crossing point at $t=0.201p$ (Fig. 1c), the coupling strength reaches its minimum, and the forward/backward GMs get close together. In this regime, due to the weak coupling between the structure and the background environment, the linewidths of both the $R_0$ dip and $R_{-1}$ peak trajectories become much smaller. As shown in the following, both BIC and EPs arise near the crossing area of the forward/backward GMs in this weak coupling scenario.

Fig. 2a shows the corresponding band structure (eigen-frequency Re($k_0$) versus parallel wavevector $k_x$) in the weak coupling regime, for the same parameters as in Fig. 1c. The coupled GMs are highlighted by red and blue curves, consistent with the dip/peak trajectories of the $R_0$/$R_{-1}$ spectra in Fig. 1c. Remarkably, by zooming close to the crossing points, we find a degenerate line around $k_x$=0.5π/$p$, as shown in Fig. 2b. Since the metagrating is an open system, the imaginary part of the eigen-frequency is generally non-zero within the light cone, leading to a finite radiative Q-factor of the GM leaky modes. Fig. 2c shows the radiative Q-factor dispersion of the coupled GM, corresponding to the modes indicated by the same color in Fig. 2b. At $k_x$=0.5π/$p$, although Re($k_0$) of the two modes is degenerate, their Q-factors (or Im($k_0$)) are largely different. One mode (mode 1) has a lower Q-factor, while the other (mode 3) has a diverging Q-factor, corresponding to a BIC. In addition, the Q-factors become degenerate when |$k_x$-0.5π/$p$| exceeds the threshold value 0.006π/$p$, at which both Re($k_0$) and Q-factor (Im($k_0$)) are degenerate (mode 2), implying the emergence of EPs within this diffractive regime. Figs. 2d-2f show field patterns for eigen-modes 1-3. For modes 1 and 2 (EP mode), the GM field penetrates the slit area (convex field distribution), and some fields leak into the background. In contrast, mode 3 (BIC mode) is characterized by fields strictly confined in the dielectric slab layer, with a null just below the slit area, forming a concave field distribution. There is no radiation to the upper semi-space, reflecting the bound nature of the mode, despite supporting a transverse momentum compatible with radiation.

Fig. 3 shows the evolution of the modes as we continuously change the coupling strength. By varying the slit layer thickness, the localized gap mode bridging energy to the background environment and the guided modes is tuned, leading to different coupling strengths between forward and backward GMs. The resonance wavelength of the localized mode can be roughly predicted by the Fabry-Perot (FP) condition, $m\lambda_m/2n_{eff}=t$, where the integer $m$ is the mode index and $n_{eff}$ is the effective index of the subwavelength slit waveguide. When the slit layer thickness $t$ is tuned such that $\lambda_m$ coincides with the GM eigen-frequency at the Brillion boundary (cases II, IV, VI in Fig. 3a, b), the strong interaction between the gap LM, the forward and backward GMs gives rise to an avoided crossing with maximum Rabi splitting. The corresponding field patterns (Fig. 3c) show that the eigenmodes are strongly coupled, and become hybrids of the gap LM (field concentrated in the slit area) and GM (field concentrated in the dielectric layer). In addition, we always find one mode with field totally confined in the dielectric layer vanishing below the slit (concave mode), due to the destructive interference of LM and GMs. Although such concave mode lies above the light line and thus within the FDO regime, it does not sustain radiation, corresponding to a BIC in the diffractive regime.

For other values of $t$, for which the LM resonance frequency deviates from the GM at the Brillouin boundary, the mutual coupling strength between forward and backward GMs decreases, leading to reduced Rabi splitting, while the BIC arises either on the upper or lower quadratic band (Supplementary Fig. S3). In particular, when $t$ is chosen such that $(m+1/2)\lambda_0/2n_{eff}=t$, the GM eigen-frequency at the Brillouin boundary satisfies $k_0=2\pi/\lambda_0=p$ (cases I, III, V), i.e., the GM is located in the middle between two adjacent ordered LMs, representing the largest separation between GM and LMs. In this situation, the coupling strength is minimum, with zero Rabi splitting with respect to the real part of eigen-frequency Re($k_0$). At the same time, the imaginary part of the eigen-frequency Im($k_0$) splits at this point, with one mode yielding zero (BIC state) and the other one yielding a finite radiation loss (Supplementary Fig. S4).

Multiple diffraction orders of the metagrating support radiative loss channels, resulting in a non-Hermitian system. For the minimum coupling (I, III, V) scenario, the LM eigen-frequencies are far away from the GMs, and the effective Hamiltonian of the system can be written as coupled GM resonances between forward and backward modes

$$H_{eff} = \begin{pmatrix} \omega_0 & v_g \left|k_x - \frac{\pi}{p}\right| \\ v_g \left|k_x - \frac{\pi}{p}\right| & \omega_0 - i\gamma_d \end{pmatrix}, \tag{1}$$

for which the complex eigenvalues are

$$\omega_\pm = \omega_0 - i\frac{\gamma_d}{2} \pm v_g \sqrt{\left|k_x - \frac{\pi}{p}\right|^2 - \left(\frac{\gamma_d}{2v_g}\right)^2}. \tag{2}$$

Here, $v_g$ is the group velocity, $\gamma_d$ is the radiative loss rate difference between convex and concave GM. This expression indicates that EPs arise at

$$k_x = \frac{\pi}{p} \pm \frac{\gamma_d}{2v_g}, \tag{3}$$

which ensures that the complex eigen-frequencies becomes degenerate. This condition is slightly away from the Brillioun boundary $k_x = \pi/p$, as numerically confirmed in the band diagrams of both Re($k_0$) and Im($k_0$) in Supplementary Fig. S4.

We can use our metagrating platform, supporting tailored BICs and/or EPs in the diffractive regime, to demonstrate sensitive wavefront shaping applications embedded in a continuous broadband spectrum. Fig. 4a shows how these features can be used for anomalous reflection. Near the BIC resonance (middle panel of Fig. 4a), anomalous reflection occurs due to near-unitary -1$^{st}$ diffraction. However, when the working frequency $k_0$ slightly deviates from the resonance condition, the impinging beam undergoes ordinary specular reflection, as shown in the upper and lower panels of Fig. 4a. The frequency difference is so small that we are not able to distinguish the different wavelength between the anomalous and ordinary reflection scenarios from the field patterns. Fig. 4b shows wavefront focusing using a quadratic phase modulation across the aperture [47-49]. Only at resonance the impinging beam can be focused at the -1$^{st}$ diffraction order (middle panel of Fig. 4b), while at other slightly detuned frequencies the impinging beam undergoes normal specular reflection (upper and lower panels of Fig. 4b).

In conclusion, here we demonstrated perfect diffraction metagratings supporting tailored BICs and EPs. By continuously tuning the coupling between the supported LM and forward/backward GMs in the FDO regime through the slit layer thickness, we observed continuous Rabi splitting evolution between coupled modes, and obtained diffractive BICs at the Brillouin boundary, as well as two diffractive EPs. The bands supporting BICs and EPs correspond to the peak trajectory of the -1$^{st}$ diffraction efficiency spectra, providing a direct link between the emergence of BICs and EPs and the diffraction response of the underlying metagratings. With the sharp spectral features of the BIC, the metagratings offer exciting opportunities for novel spectrum sensitive diffraction and wavefront shaping applications. Leveraging EPs in the diffractive regime, robust wavefront phase modulation with topological features can be envisaged [50]. Our findings may find interesting implications for embedded wavefront shaping and sensing applications.

This work was supported by the National Natural Science Foundation of China (NSFC) (Grants 62075084), the National Key R&D Program of China (2018YFB1107200), the Guangdong Provincial Innovation and Entrepreneurship Project (Grant 2016ZT06D081), the Guangdong Basic and Applied Basic Research Foundation (2020A1515010615), the Guangzhou Science and Technology Program (202102020566), the Fundamental Research Funds for the Central Universities (21620415), the Simons Foundation and the Air Force Office of Scientific Research MURI program.

**Figures**

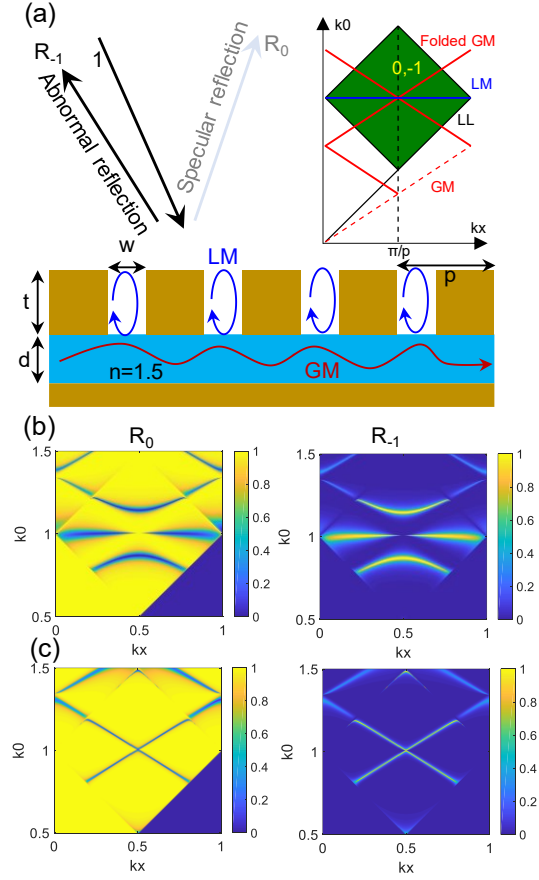

FIG.1. (a) Schematic of our metagrating geomertry, composed of a metallic subwavelength slit array with period $p$, slit width $w$ and thickness $t$, on top of a dielectric spacer with refractive index $n$=1.5 and thickness $d$, and a metallic substrate. Localized (LM) and guided (GM) modes emerge in the metallic slit and dielectric spacer layer, respectively, providing a powerful way to manipulate and control the diffraction orders $R_0$ and $R_{-1}$, within the few-diffraction-order regime, as denoted by the green area of the dispersion diagram in the inset. (b, c) Diffraction efficiency of $0^{th}$ (left panels) and $-1^{st}$ (right panels) orders as we vary the parallel wavevector $k_x$ and total wavevector $k_0$, for different slit layer thicknesses: (b) $t$=0.425$p$ corresponding to strong coupling and (c) $t$=0.201$p$ corresponding to weak coupling. The metal is treated here as a perfect electric conductor (PEC), which is a good approximation for metal at terahertz and microwave frequencies.

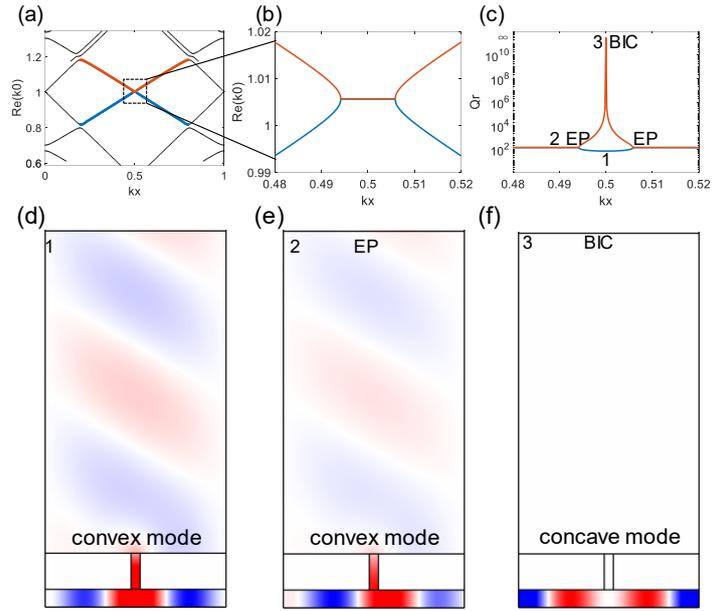

FIG.2. Band diagram and field patterns of eigenmodes in the anti-crossing region, with same parameters as in Fig. 1c. (a) Band diagram of the real part of $k_0$ versus $k_x$ and its (b) zoomed region near the crossing point. (c) Radiative quality (Q) factor. (d-e) Field patterns ($H_z$) of 1. lowest Q-factor radiative mode, 2. EP mode, and 3. BIC mode as indicated in (c).

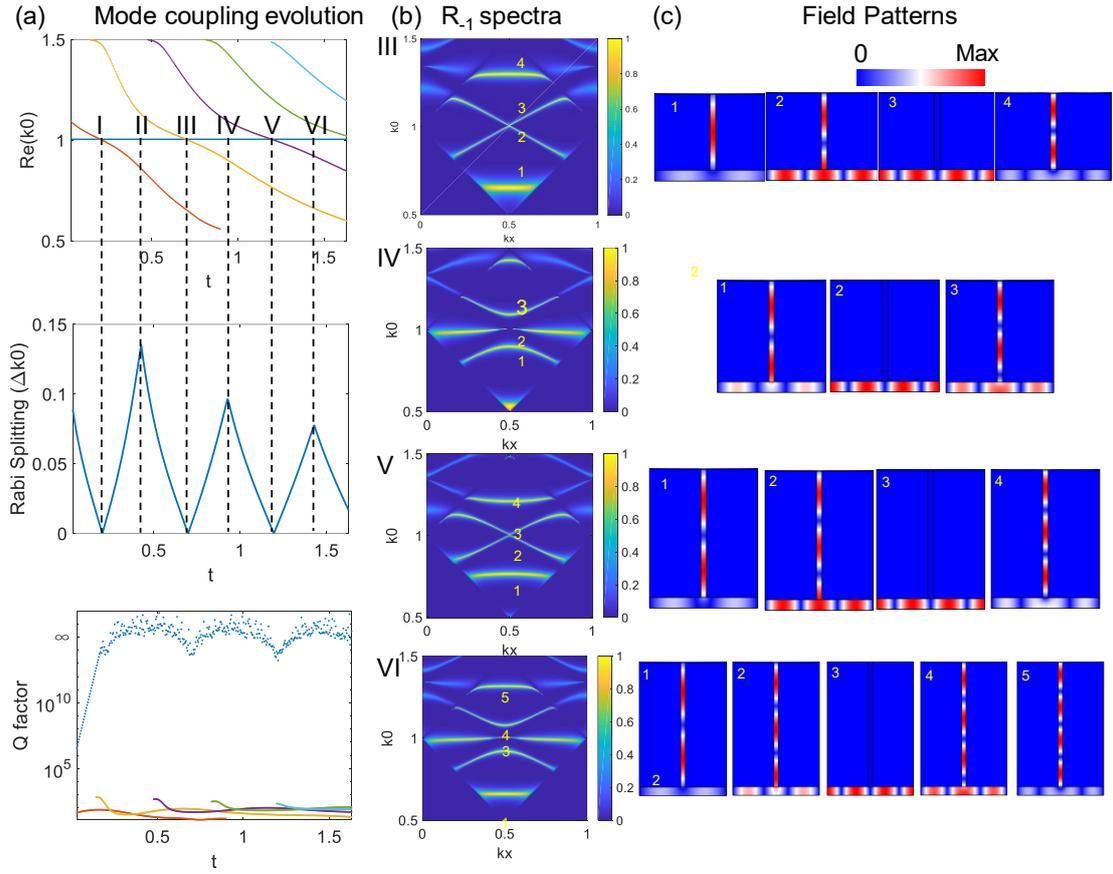

FIG. 3. Evolution of mode coupling by varying the slit layer thickness. (a) Upper panel: eigen-frequencies, middle panel: Rabi splitting, and lower panel: radiative Q-factor evolution with slit layer thickness $t$. (b) -1st diffraction efficiency spectra with varying parallel wavevector $k_x$ and total wavevector $k_0$, for higher-order weak coupling cases with (III) $t=0.698p$, (V) $t=1.194p$, and strong coupling cases with slit layer thickness (IV) $t=0.946p$, (VI) $t=1.442p$, respectively. The other parameters are $w=0.5p$, $d=0.1p$ fixed. (c) Field patterns corresponding to the modes indicated in each band of (b).

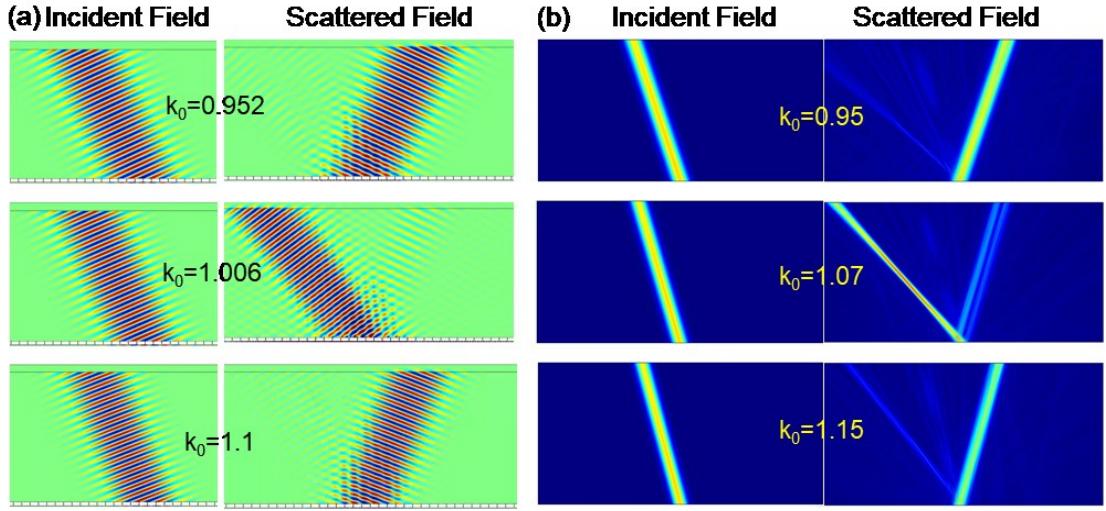

FIG. 4. Embedded wavefront shaping. (a) Field patterns ($H_z$) demonstrating wavefront deflection of a Gaussian beam incident on a periodic metagrating (with parameters $w=0.05p$, $t=0.425p$, $d=0.1p$) tuned at the quasi-BIC resonance (middle panel) and a little detuned from the quasi-BIC resonance (upper and lower panels). (b) Field patterns ($|H_z|$) demonstrating wavefront focusing of a Gaussian beam incident on a modulated metagrating (with parameters $w=0.2p$, $t=0.425p$, $d=0.1p$) with quadratic phase profile at the quasi-BIC resonance (middle panel). When the wavelengths are slightly shifted from the quasi-BIC frequency, a focusing phenomenon in the abnormal reflection direction is replaced by specular reflections (upper and lower panels).

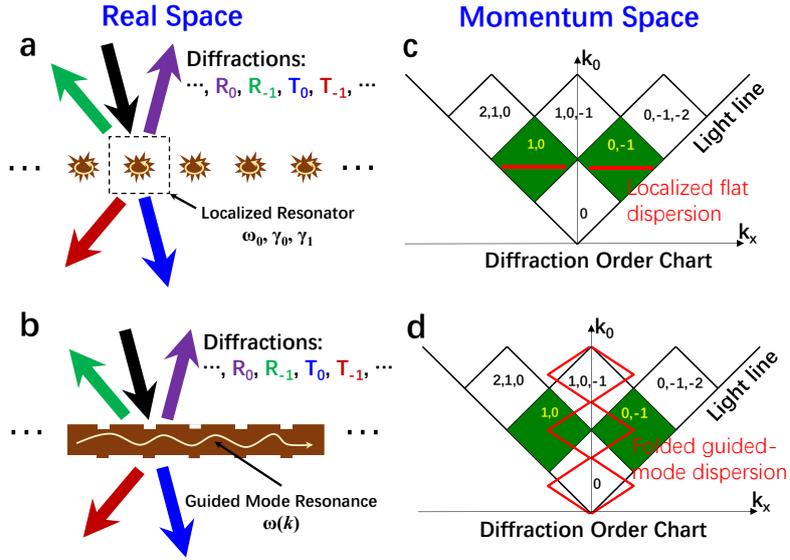

FIG. S1. Schematic of diffraction order management for (a, c) localized mode and (b, d) guided mode mediated metagrating, in (a, b) real space and (c, d) momentum space. Localized mode mediated metagrating relies on localized resonances of the inclusion for diffraction order management (a), which usually manifest themselves with a flat near-unitary diffraction efficiency trajectory in the few-diffraction-order regimes, where there are only $0^{th}$ and $-1^{st}$ or $1^{st}$ orders (denoted as green areas), as shown in the diffraction order chart in (c). Guided mode mediated metagratings use dispersive guided mode resonances to manage multiple diffraction orders (b). In the diffraction order chart (d), the dispersion curves of the guided mode are folded into multiple diffraction regimes. In the few-diffraction-order regime, the trajectories of guided mode resonances lead to angularly dispersive near-unitary diffraction efficiencies.

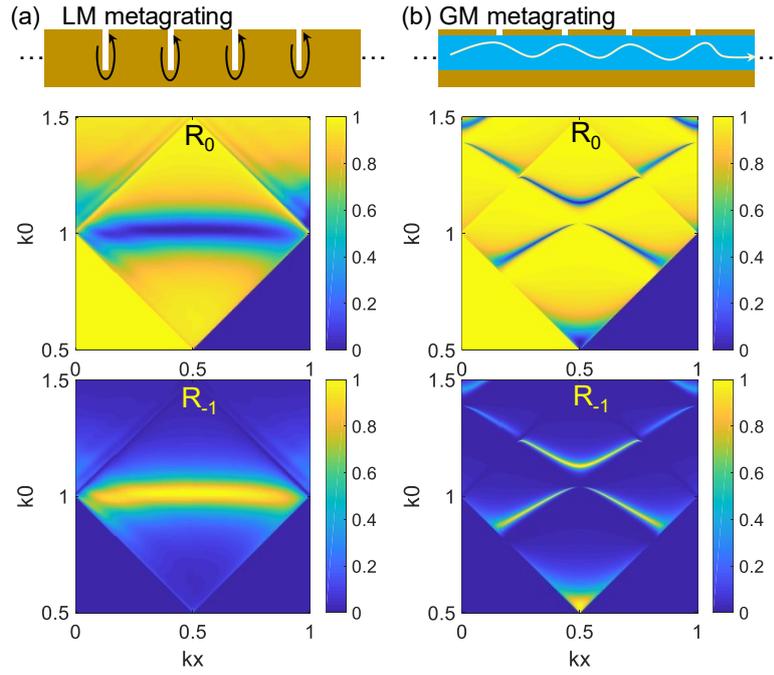

FIG. S2. Comparison of purely localized mode and purely guided mode mediated metagratings. (a) Schematic structure (upper panel), $R_0$ diffraction spectra (middle panel) and $R_{-1}$ diffraction spectra (lower panel) of a pure localized mode mediated metagrating implemented by the metallic groove array. (b) Schematic structure (upper panel), $R_0$ diffraction spectra (middle panel) and $R_{-1}$ diffraction spectra (lower panel) of a pure localized mode mediated metagrating implemented by a metal-dielectric-metal waveguide incorporated with a thin layer of metallic slits.

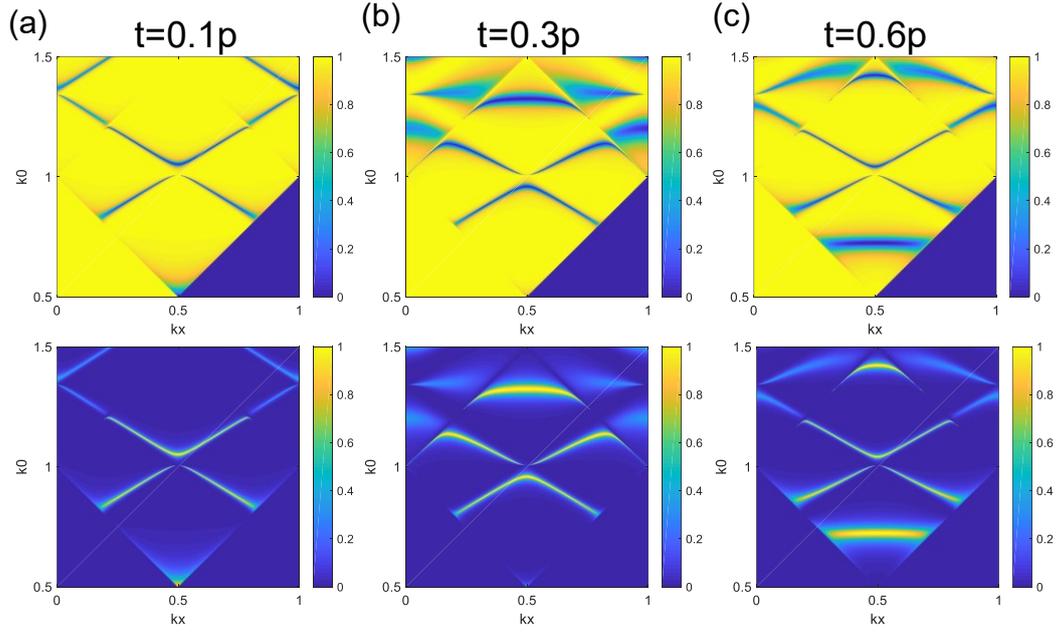

FIG. S3. Diffraction spectra with varying $k_x$ and $k_0$ of the metagrating of moderate coupling strengths at (a) $t=0.1p$, (b) $t=0.3p$, (c) $t=0.6p$, respectively. The upper panels are specular reflection ($R_0$) spectra, and the lower panels are -1st order reflection spectra ($R_{-1}$). At these moderate coupling strengths between the maximum and minimum coupling strengths, the BIC point always exists as indicated by the linewidth vanishing point at the Brillouin boundary $k_x=0.5\pi/p$. The BIC may lie on the lower quadratic band (a, c), the upper quadratic band (b), or the flat band (Fig. 1b).

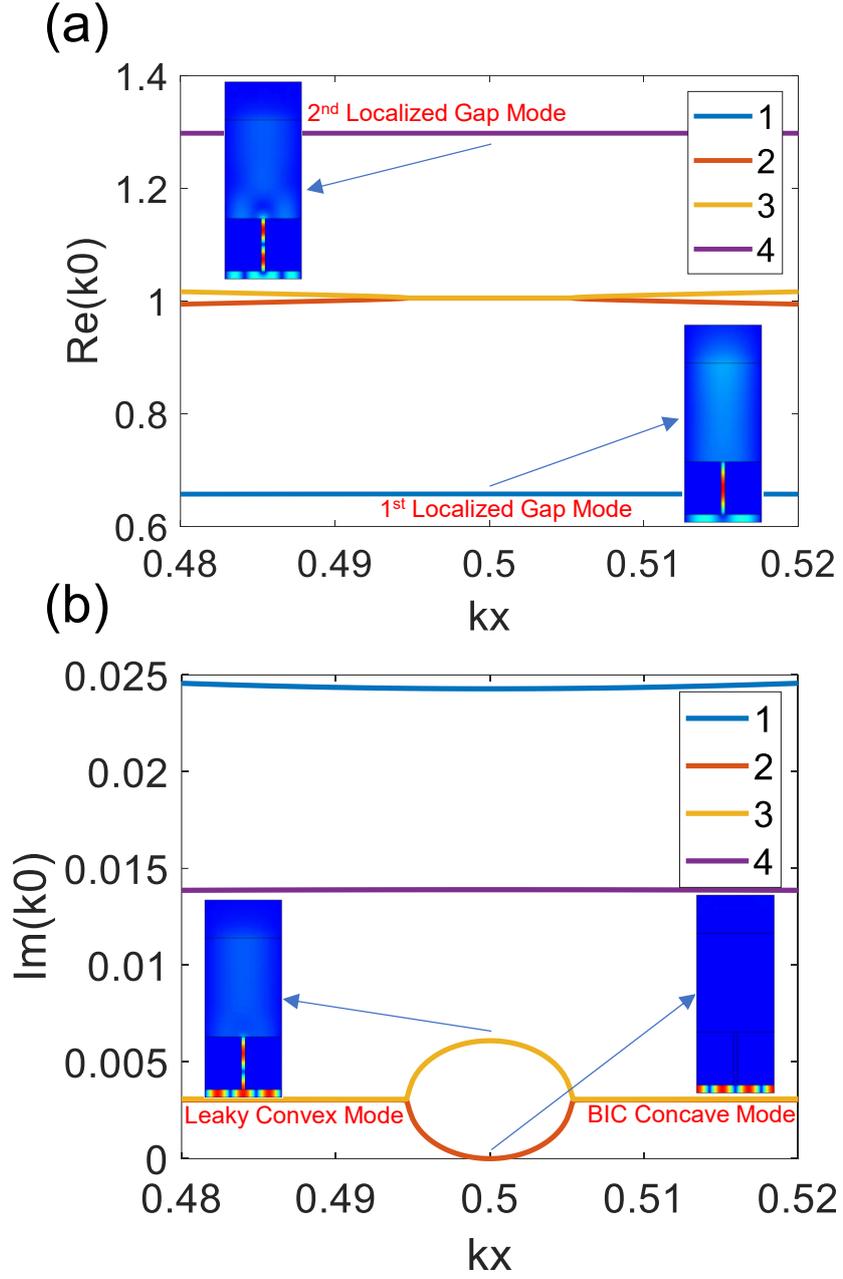

FIG. S4. Band diagrams for both (a) Re($k_0$) and (b) Im($k_0$) of a higher-order weak coupling metagrating with $w=50p$, $t=698p$. BIC and EP states co-exist resulting from the interaction between higher-mode LMs in the vertical metallic slits with GMs in the horizontal MIM waveguide, leading to leaky convex and concave modes as shown in the insets.

4